\documentstyle[12pt]{article}
\def\hq{{\hat q}}
\def\hp{{\hat p}}
\def\hA{{\hat A}}
\def\hB{{\hat B}}
\def\hS{{\hat S}}
\def\hL{{\hat L}}
\def\hH{{\hat H}}

\title{ BRST-BFV quantization and the  Schwinger action
principle}

\author{ J. ANTONIO GARCIA and  J. DAVID VERGARA \\
Instituto de Ciencias Nucleares \\
Universidad Nacional Aut\'onoma de M\'exico \\
Apartado Postal 70-543, 04510
M\'exico, D.F. \\
and \\
LUIS F. URRUTIA \thanks{On sabbatical leave from Instituto de Ciencias
Nucleares, Universidad Nacional Aut\'onoma de M\'exico,
Circuito Exterior, C.U., 04510 M\'exico, D.F.}  \\
Departamento de F\'\i sica \\
Universidad Aut\'onoma Metropolitana-Iztapalapa \\
Apartado Postal 55-534, 09340 M\'exico, D.F. \\
and  \\
Centro de Estudios Cient\'\i ficos  de Santiago \\
Casilla 16443, Santiago 9 , Chile.}

\begin{document}
\maketitle

\begin{abstract}

We introduce an operator version of the BRST-BFV effective action for
arbitrary systems with first-class constraints.
Using the Schwinger action principle
we calculate the propagators corresponding to:
(i) the  parametrized non-relativistic free particle,
(ii) the  relativistic free particle and (iii)
the spining relativistic free particle.
Our calculation correctly imposes the  BRST-invariance at the end-points.
The precise use of the additional boundary terms required in the description
of fermionic variables is also incorporated.

\noindent
(PACS:  04.20.Fy  \ \ 04.60. Ds \ \ 11.10.Ef     )
\end{abstract}

\baselineskip=20pt

\section{Introduction}

A wide variety of  interesting  theories in physics  which  range, for
example,
from the standard model of strong, weak and electromagnetic interactions, to
Einstein general relativity and even to  more speculative ideas like string
theories, can be understood and unified  under the generic label of
constrained
systems. All such gauge theories
are characterized by  the existence of relations (
constraints) among the original phase space variables, together with the
appearance of arbitrary functions
in the solutions of the equations of motion. The quantum mechanical
description
of such systems
deviates from the standard prescriptions, like the canonical or path integral
quantization for
unconstrained systems. A systematic procedure for dealing with the
quantization
of constrained systems was proposed some time ago  by Dirac \cite{dirac} and
recently the method  has been  extended to the BRST-BFV prescription
\cite{fradkin} and also to the antifield method of Batalin-Vilkovisky
\cite{batalin},
both  in the context of  the path integral and operator
approach to quantization. These methods have been successfully  applied to
many different problems such as supergravity \cite{kal} , topological field
theories \cite{bir}  and superstrings \cite{hat}, just to mention a few
interesting cases.

Nevertheless, in the literature
we can find alternative methods of quantization, among which  the  Schwinger
action principle
\cite{schw} constitutes a very important case. This action principle  can be
applied to
arbitrary quantum systems and starts from an operator formulation
of the action from the very
 beginning. The general validity of this principle  has recently motivated  its
application to  the case of constrained
systems. For example, in reference \cite{das} it is shown that when  the
Schwinger action principle is  applied to a system with only  second class
constraints, it  leads to (anti)commutations relations
corresponding precisely
to  the Dirac bracket prescription. Another application of this
action principle
has been the calculation of the
quantum-mechanical  BRST-invariant matrix elements
of the evolution operator in the cases of   the spinless and the spining
relativistic free particle
\cite{rabello}, which were previously  obtained
using the
BRST-BFV path integral formulation in Refs. \cite{henneaux},\cite{batlle}.
Unfortunately, the  calculation in Ref. \cite{rabello} makes  use of an
incorrect
(i.e. non-BRST invariant) basis for the physical states at the initial and
final times.

The Schwinger action  principle can be viewed as a generalization
of the Weiss action principle in classical mechanics to the
quantum case . The Weiss principle states that if we make
an arbitrary
variation of the action
\begin{equation}
S=\int^{t''}_{t'}L(\dot q, q, t)\, dt,
\end{equation}
the Euler-Lagrange equations follows from the requirement
\begin{equation}
\delta S= G(t'')-G(t'),
\end{equation}
where the $G$'s are boundary terms  \cite{sudarshan} . This means that such
variation
must not depend upon the  trajectory that connects the
end-points, thus leading to the Euler-Lagrange  equations of motion.
The explicit  form of the
boundary terms
depends upon the dynamical variables which are kept fixed at boundaries.
For example, if we fix the coordinates $q$ at the boundaries  $t''$ and $t'$,
then
\begin{equation}
G(t)=(p\delta q- H\delta t)|_{t },
\end{equation}
where $H$ is the corresponding Hamiltonian function and the notation is
$p\delta q=
p_i \delta q^i$,  with  the summation being over all  the degrees of freedom
of the system. The notation of deleting the indices will be frequently
used in the sequel, in the cases where no confusion arises.

At the quantum level, the Weiss action principle is replaced by the
Schw\-in\-ger
action principle which states that
the arbitrary variation of the matrix
elements of the
evolution operator  $<a''| U(t'',t') |b'>
\equiv \langle a'' t''|b' t'\rangle$ is given by
\begin{equation}
\delta \langle a'' t''|b' t'\rangle=i\langle a'' t''|\delta\bigg(
\int^{t''}_{t'}L(\dot{\hq}, {\hq}, t)\, dt\bigg)|b' t'\rangle. \label{delamp}
\end{equation}
Moreover, the variation of the hermitian action operator
$\hS=\int^{t''}_{t'}L(\dot \hq, \hq, t) dt$
must depend only
upon the
end point operators and times, in such a way that
\begin{equation}
\delta\bigg(\int^{t''}_{t'}L(\dot \hq, \hq, t)\, dt\bigg)=G(\hA'',
t'')-G(\hB', t'), \label{delL}
\end{equation}
where $\hA''$ denotes a complete set of commuting operators at  the time $t''$
with corresponding eigenvalues $a''$ and analogously for the operators $\hB'$
at $t'$.
That is to say, the variation of the propagator is given by the corresponding
matrix elements of the variation of a single
quantum mechanical operator:
the action operator.

A convenient choice for the quantum  Lagrangian is the  first order form $\hL=
{1 \over 2}\left( \hp \dot \hq + \dot \hq \hp  \right)- H( \hq, \hp, t ) $,
where
 $H(\hq, \hp, t ) $ is the hermitian  Hamiltonian operator constructed in the
usual  way
starting from  the definition $\hp={\partial \hL \over \partial \dot \hq}$.
The resulting equations of motion are the standard Hamilton equations  for the
quantum operators
\begin{equation}
\dot \hq={\partial \hH \over \partial \hp},  \qquad  \dot \hp=-{\partial \hH
\over \partial \hq}\label{hei}
\end{equation}
 and one can identify the  corresponding  end-point generators as
\begin{equation}
G(\hat q, \hat p) =\hp \delta \hq -\hH \delta t  \equiv G_{{\delta }_{\hq}} +
G_{{\delta}_{t}}.
\end{equation}

At this stage it  is  also necessary  to  specify the operator  character of
the
variations $\delta \hq, \delta \hp$
which imply the above results. When dealing with
bosonic (fermionic) operators, called operators of the first (second)
kind in Schwinger's notation, the corresponding variations  satisfy the
standard commutation (anticommutation) rules of even (odd) elements
in a Grassmann algebra. The standard notation is that bosonic(fermionic)
objects have even(odd) Grassmann parity in the underlying Grassmann algebra.

By considering a canonical transformation which interchanges the roles of
$\hq$
and $\hp$, it is possible to identify the generator of  infinitesimal
transformations in $\hp$ as $G_{{\delta}_{ \hp}}= -\hq \delta \hp$. The above
expressions for the generators of the corresponding transformations, together
with the quantum mechanical interpretation of them as
producing infinitesimal unitary
transformations,  leads to the general commutator
\begin{equation}
\left[ \hat A, \  {\hat G}_{{\delta }_{{\hat b}}}  \right]=i
\delta_{\hat b}(\hat A).\label{cr1}
\end{equation}
{}From this expression we obtain  the basic (anti)commutation relations  for
the phase space
variables, after   taking  into account the (anti)commutation properties
of the parameters associated to the above transformations.

{}From Eqs. (\ref{delamp}) and (\ref{delL}), the final expression for arbitrary
variations of the
propagator is then given by
\begin{equation}
\delta\langle a'' t''|b' t'\rangle=i \langle a'' t''|G(A'', t'')-G(B', t')|b'
t'\rangle.\label{acp}
\end{equation}

In order to use the above expression as a
practical computational tool,
we must be able to solve the operator Heisenberg equations of motion
for the system in terms of the operators $A'', B' $ whose eigenvalues are
kept fixed at
the end-points. In this way, we will be  able to calculate the corresponding
matrix  elements in (\ref{acp}),
which provide a set of partial differential equations for the propagator, that
must be subsequently integrated.
In other words, we need to choose a complete set of  (anti)commuting
operators at the initial and final times, together with a well
defined inner product in the Hilbert space of  physical states,
in order to specify the quantum numbers at
the end-points which, of course,  must be compatible with the
dynamics.

In this paper we introduce  an operator BRST-BFV action  for arbitrary systems
with first-class
constraints, which is inspired in Schwinger action principle. This action is
defined with
appropriate BRST-invariant boundary conditions.
As an application of this quantum action  and  as
an alternative procedure to the standard path-integral approach,  we carefully
calculate the
propagators corresponding to the non-relativistic particle,
the relativistic spinless particle and the relativistic
spining particle. The corresponding calculations using  the BRST-BFV
path integral approach can be found in  Refs.\cite{henneaux}, \cite{batlle},
 \cite{gomis2}.
The results presented in our work are based on a consistent  choice of
end-points conditions and thus allows to clarify some incorrect points that
arise in Ref. \cite{rabello}.

The paper is organized as follows: section 2 contains our general prescription
to construct
the quantum BRST-BFV action, from which we subsequently calculate the
corresponding  propagators  using the Schwinger action principle. The
next  sections, 3, 4 and 5,  contain
the corresponding  calculations for the following  particular cases : the
parametrized non-relativistic particle, the relativistic free particle and the
spining relativistic free particle, respectively.

\section{The  quantum BRST- BFV action}

Since the action principle does not provide a quantum action to start with,
we follow  the usual procedure of defining the
quantum action as a consistent  extension of
the classical action associated to the problem.

For a system with constraints, one of the most successful prescriptions to
construct a
classical gauge independent action is the BRST-BFV method \cite{libro}. The
resulting  action has the
advantage of being invariant under BRST transformations and since
the remaining symmetry is only global, all the variations of the
canonical variables are independent.

We start from a classical
system described by  canonical coordinates $q^i, p_i \ \\
(i=1,\dots,n)$,
having  only first-class constraints $G_a(p,q) \ (a=1,\dots , m)$, and with a
first-class canonical Hamiltonian $H_0(q,p)$
\begin{equation}
G_a(q,p)\approx 0
\end{equation}
\begin{equation}
\{G_a, G_b\}_{PB} = C_{ab}^{\ \ c} (q,p) G_c, \quad
\{G_a, H_0\}_{PB} = D_{a}^{\ \ c} (q,p) G_c.
\end{equation}
We assume, for simplicity, that all second-class
constraints have been
eliminated, either by solving them or by transforming them into
first-class constraints, adding
new variables and
using, for example, the Batalin-Tyutin conversional method \cite{baty}.

Consider the variational principle in the class of paths
$q^i(\tau)$, $p_i(\tau)$, $\lambda^a(\tau)$, where $\lambda^a(\tau)$
are Lagrange multipliers associated to the constraints, with
prescribed values
at the
endpoints $\tau'$ and $\tau''$,
\begin{equation}
Q_i(q(\tau'),\ p(\tau'), \tau')={Q'}_i, \qquad
Q_i(q(\tau''),\ p(\tau''), \tau'')={Q''}_i, \label{boun}
\end{equation}
of a complete set of commuting variables $Q_i(q,p,\tau)$
 \begin{equation}
\{Q_i , Q_j\}_{PB} =0,  \  \ \hbox{(at equal times)}.
\end{equation}
The action for this variational principle is
 \begin{equation}
S[q^i(\tau),p_i(\tau),\lambda^a(\tau)]=
\int_{\tau'}^{\tau''} (\dot q^i p_i  - H_0 -\lambda^a G_a) d\tau -
B(\tau'') + B(\tau'), \label{classac}
\end{equation}
(for paths obeying  (\ref{boun})), where the phase space function
$B(q,p,\tau)$ is such that
\begin{equation}
p_i \delta q^i =- P^i \delta Q_i + \delta B,
\end{equation}
for fixed $\tau$ \cite{hd}. Here, the $P^i$ are the momenta canonically
conjugated to the
$Q_j$,
\begin{equation}
\{P^i ,  P^j\}_{PB} =0, \ \ \ \ \{Q_i , P^j\}_{PB}=\delta_i^j.
\end{equation}
We assume that (\ref{classac}) is  the final action of the
system arising  after we have completed the Dirac procedure of generating all
possible secondary constraints and after we have  eliminated  all  second-class
constraints. This means that  we have already enforced the consistency
conditions $\dot G_a \approx 0$.

In order to construct the BRST-BFV effective action according to Ref.
\cite{libro}, we start from a configuration space where all degrees of freedom,
which can have either even or odd Grassmann parity,
are real.  Also,  we choose the Lagrangian to be real and even.  If some
coordinate $\theta$ is fermionic ( odd Grassmann parity),   the corresponding
momentum $p_\theta$ is imaginary and odd in such a way that
${\dot\theta} p_\theta$ is real an even. We assume also that
all the constraints are real. They can have  either odd or even  Grassmann
parity. In the later
case $\lambda^a$ is imaginary and odd, so that $\lambda^aG_a$ is both real
and even. Next we
promote the Lagrange multipliers to the status of
dynamical variables by introducing their corresponding canonically conjugated
momenta $\pi_a$ and we demand that $\pi_a \approx 0$, in such a
way that we have now $2n$ first-class constraints $G_A=(\pi_a, G_a)\approx 0$.
The Grassmann parity $\epsilon$ of the new variables is such that
$\epsilon(\pi_a)=\epsilon(\lambda^a)
=\epsilon (G_a)\equiv \epsilon_a$.  The next step is to introduce the
ghost variables  $ \eta^A$ together with the corresponding  anti-ghost
variables $ {\cal P}_A$,  in such a way that
$ \epsilon(\eta^A)= \epsilon({\cal P}_A)= \epsilon_a +1.$  Following the
standard convention we consider the splitting
\begin{equation}
\eta^A=(-(i)^{\epsilon_a+1}{\cal P}^a,{\cal C}^a),
\quad {\cal P}_A=((i)^{\epsilon_a+1}{\bar {\cal C}}_a,{\bar
{\cal
P}}_a).
\end{equation}

The classical  effective BRST-BFV  action turns out to be
\begin{equation}
S_{BRST}=\int_{\tau'}^{\tau''}\left({\dot q}^ip_i-
\lambda^a{\dot\pi}_a +
{\dot{\bar {\cal C}}}_a{\cal P}^a+{\dot{\cal C}}^a{\bar {\cal
P}}_a - H_{BRST} \right)d\tau - [B]_{\tau'}^{\tau''}, \label{effac1}
\end{equation}
where the integral is extended over the paths which obey the boundary
conditions
$$
Q_i(q(\tau'),\ p(\tau'), \tau')=Q'_i, \qquad
Q_i(q(\tau''),\ p(\tau''), \tau'')=Q''_i.
$$
\begin{equation}
{\cal C}^a(\tau') = {\cal C}^a(\tau'') =0, \qquad
{\bar {\cal C}}_a(\tau') = {\bar {\cal C}}_a(\tau'') =0, \label{bc1}
\end{equation}
$$
{\pi}_a(\tau') = {\pi}_a(\tau'') =0.
$$
In Eq.(\ref{effac1}),  $H_{BRST}=H_c-\{\Psi,\Omega\}_{PB}$, $H_c$ is
the canonical Hamiltonian,  $\Psi$ is the so
called
fermionic gauge-fixing term and
$\Omega$ is the nilpotent BRST-charge, which has odd Grassmann parity and
satisfies
 $ \{ \Omega, \ \Omega \}_{PB}=0 $.  The general form of the BRST charge is
$\Omega= -(i)^{\epsilon_a +1} {\cal P}^a \pi_a + {\cal C}^a G_a +$
 ``more'' , where ``more'' stands for terms at least quadratic in the
ghosts. A systematic algorithm for
this construction can be found in Ref. \cite{libro}.
In all the applications that  we will consider in this work, we
choose the  classical  fermionic gauge to be
\begin{equation}
\Psi={\bar {\cal P}}_a \lambda^a, \label{FG1}
\end{equation}
which has odd Grassmann parity.

Let us  observe that we can read out directly from the action (\ref{effac1})
the classical Poisson brackets for the fundamental variables
$$\{p_i,q^j\}_{PB}=-\delta^j_i=(-)^{\epsilon(q^j)+1}\{q^j,p_i\}_{PB} \ ,$$
$$\{\pi_a,\lambda^b\}_{PB}=-\delta^b_a=(-)^{\epsilon_a+1}
\{\lambda^b,\pi_a\}_{PB} \ ,$$
$$\{{\cal P}^a,{\bar{\cal C}}_b\}_{PB}=-\delta^a_b=(-)^{\epsilon_a}
\{{\bar{\cal C}}_b,{\cal P}^a\}_{PB} \ ,$$
$$\{{\bar{\cal P}}_a,{{\cal C}}^b\}_{PB}=-\delta_a^b=(-)^{\epsilon_a}
\{{{\cal C}}^b,{\bar{\cal P}}_a\}_{PB} \ .$$

The above action (\ref{effac1}) has two important
properties: (i)  all canonical variables involved  are unconstrained.
This feature is  reflected  in the choice of  the associated
measure in
the path integral formulation of the method,
as  the corresponding  Liouville measure.
(ii) the remaining symmetry of the action
(\ref{effac1}) is a global supersymmetry generated by the
BRST charge $\Omega$, which imposes the choice of BRST-invariant end-point
conditions.

The classical
effective action (\ref{effac1}) is our starting point to construct
the quantum version of the BRST-BFV
method.
First, we promote all (imaginary)real phase space variables $A$, including the
ghosts, to
(antihermitian)her\-mitian operators ${\hat A}$.
Since the quantum action must be hermitian in order to preserve unitarity,
we also  adopt the standard replacement for extending real classical products
of
real variables into hermitian products of hermitian quantum operators
\begin{equation}
(i)^{\epsilon(A)\epsilon(B)}AB\rightarrow{1\over2}(i)^{\epsilon(A)
\epsilon(B)}({\hat A}{\hat B} + (-1)^{\epsilon(A)\epsilon(B)}{\hat B}{\hat
A})
\equiv <<{\hat A}{\hat B}>>.\label{simm}
\end{equation}
Let us observe that the operator properties  assumed for the variations
$\delta
{\hat A}, \delta{\hat  B}$  precisely
guarantee that  $ \delta <<{\hat A}{\hat B}>>=(\delta {\hat A}){\hat B}+ {\hat
A} (\delta
{\hat B})$. In particular,
the above prescription has to be applied both to the kinetic term and to the
boundary
term  in the action. The quantum expression for the
latter will be discussed in each separate situation
and the specific form
will be  dictated by the classical counterpart.
The interplay among the
variations of
both types of terms will allow the proper identification of the corresponding
quantum generators at the end points,
thus providing the basic (anti)commutation relations
for the dynamical variables directly from the  action
principle.

According to the above prescription,
the quantum extension of the BRST charge
must lead to
an hermitian operator ${\hat \Omega}^\dagger ={\hat \Omega}$ such that  $\{
{\hat \Omega}, \  {\hat \Omega} \} = 2 {\hat \Omega}^2= 0$,
with  $\{ {\hat A}, \  {\hat B} \}={\hat A}{\hat  B} + {\hat  B} {\hat A}$
denoting  the corresponding
anticommutator.   Both, the canonical Hamiltonian together with the  fermionic
gauge fixing term
are also promoted to the corresponding  operators ${\hat H}_c$ (hermitian) and
  ${\hat \Psi}$ (antihermitian)  respectively, while the effective Hamiltonian
operator is defined by
${\hat H}_{BRST}={\hat H}_c+ i\{ {\hat \Psi }, \  {\hat  \Omega }\}$.
Besides,
the BRST charge must
be conserved i.e. $[{\hat \Omega}, \ {\hat H}_{BRST}]=0$, where
  $[ {\hat A}, \  {\hat  B}]={\hat A}{\hat  B} - {\hat  B} {\hat A}$
denotes  the corresponding commutator.

In this way, the full quantum action turns out to be
\begin{equation}
{\hat S}_{BRST}=\int_{\tau'}^{\tau''}\left( <<{\dot{\hat  q}}^i{\hat p}_i -
{{\hat \lambda}}^a{\dot {\hat \pi}}_a +
{\dot{\hat {\bar {\cal C}}}}_a{{\hat {\cal P}}}^a+{\dot{\hat{\cal
C}^a}}{\hat{\bar {\cal
P}}}_a  >> - {\hat H}_{BRST} \right)d\tau -
[\hat{B}]^{\tau''}_{\tau'}.  \label{qeffac}
\end{equation}

The basis vectors of the Hilbert space
at the initial time $\tau'$,
$|  Q'_i, {\cal C}^{\prime
a}, \bar {\cal C}^{\prime}_{ a}, {\pi}'_a \rangle$,
are labeled by the
corresponding fixed eigenvalues and satisfy
\begin{equation}
{\hat Q}_i |  Q'_i, {\cal C}^{\prime a},
\bar {\cal C}^{\prime}_{ a}, {\pi}'_a \rangle =
Q'_i
|  Q'_i, {\cal C}^{\prime
a}, \bar {\cal C}^{\prime}_{ a}, {\pi}'_a \rangle
\end{equation}
\begin{equation}
{\hat {\cal C}}^a |  Q'_i, {\cal C}^{\prime a},
\bar {\cal C}^{\prime}_{ a}, {\pi}'_a \rangle =
\hat {\bar {\cal C}}_a |  Q'_i, {\cal C}^{\prime a},
\bar {\cal C}^{\prime}_{ a}, {\pi}'_a \rangle =
{\hat {\pi}}_a |  Q'_i, {\cal C}^{\prime a},
\bar {\cal C}^{\prime}_{ a}, {\pi}'_a \rangle =0,\label{cbc0}
\end{equation}
according to the classical boundary conditions (\ref{bc1}). Analogous
expressions are valid
for the  basis vectors at the final time $\tau''$.

The invariance of the action under quantum BRST transformations
is stated in the
property
$ \delta_{\Omega} {\hat S}_{BRST}=i [{\hat \Omega}, \ {\hat S}_{BRST}]_{-}
=0 $.
The BRST invariance of the related transition amplitudes
$\langle a'' t'' | \hat S | b' t' \rangle$ is guaranteed
provided the end point states are also  invariant under this
transformation, which means that  ${\hat \Omega} | b' t' \rangle=0= {\hat
\Omega} | a'' t'' \rangle$.

\section{The parametrized non-relativistic particle}

The  classical action for this system is
\begin{equation}
S=\int_{\tau'}^{\tau''} L d\tau= \frac{m}{2}\int_{\tau'}^{\tau''}\frac{{\dot
x}^2}{{\dot t}}\, d\tau.
\end{equation}
Next we define
$p_x = {\partial L \over \partial \dot x}$, $p_t={\partial L \over
\partial \dot t}$ as the momenta canonically conjugated to the coordinates
$x$ and $t$ respectively. Here the dot means the derivative with respect to
the parameter $\tau$. In this case, the canonical Hamiltonian $H_c$
is zero and the application of
the standard Dirac procedure leads to only one (first-class) constraint
\begin{equation}
G=H_0+p_t\approx 0, \label{cons1}
\end{equation}
where
\begin{equation}
H_0\equiv\frac{p_x^2}{2m}.
\end{equation}

Our application of the Schwinger action principle will start from the
effective action operator constructed according to the ideas of the previous
section. In this case, the subindex $a$ takes just one value, corresponding
to the only constraint of the problem . The
quantum effective action is
taken to be
\begin{eqnarray}
{\hat S}_{BRST}&=&\int_{\tau'}^{\tau''}\left( <<{\dot{\hat x}}{\hat p}_x
+{\dot{\hat t}}{\hat p}_t-
{{\hat \lambda}}{\dot {\hat \pi}} +
{\dot{\hat {\bar {\cal C}}}}{{\hat {\cal P}}}+{\dot{\hat{\cal
C}}}{\hat{\bar {\cal
P}}} >> - {\hat H}_{BRST} \right)d\tau \nonumber \\
& & + <<{\hat x}'{{\hat p}_x}'
+{\hat t}'{{\hat p}_t}'>>,
\label{qacfp0}
\end{eqnarray}
where
\begin{equation}
{\hat H}_{BRST}=i \{ {\hat \Psi }, \  {\hat  \Omega } \},\quad
{\hat \Psi}= {\hat {\bar {\cal P}}}{\hat \lambda}, \quad
{\hat \Omega}=-i{\hat {\cal P}}{\hat \pi} + {\hat {\cal C}} \left(
\frac{{\hat p}_x^2}{2m}+{\hat p}_t \right). \label{addop1}
\end{equation}
In the sequel, all the canonical variables are considered to be
operators and we drop the hat on top of them in order to simplify
the notation.
The application of the action principle to the action (\ref{qacfp0}) leads
to the Heisenberg equations of motion, written in the general form of
Eqs.(\ref{hei})
in terms of the BRST-Hamiltonian,
together with the following identification of the generators
of transformations at the end-points
\begin{eqnarray}
\delta {\hat S}_{BRST}&=&\left( p_x''\delta x''+x'\delta p_x'
+p''_t\delta t''+t' \delta p_t' -\lambda''\delta\pi''
+\lambda'\delta \pi' -{\bar {{\cal P}''}} \delta {\cal C}''   \right.
\nonumber
\\
& & \left.+ {\bar {{\cal P}'}}
\delta {\cal C}'
-{\cal P}'' \delta {\bar {{\cal C}''}}+{\cal P}' \delta {\bar {{\cal C}'}}-
{H}''_{BRST}\delta \tau''
+{H}'_{BRST}\delta \tau' \right), \label{endpoint1}
\end{eqnarray}
where the superscript $'$ ($''$) denotes the evaluation of the
corresponding operator at $\tau=\tau' (\tau=\tau'')$ respectively.
According to the property (\ref{cr1}), the equation (\ref{endpoint1}) implies
the
following non-zero (anti)commutation
relations
at equal times
\begin{equation}
[x,p_x]=[t,p_t]= [\lambda,\pi]=i\qquad \{{\bar {\cal C}},{\cal P}\}=\{{\bar
{\cal P}},{\cal C}\}=-i.\label{com1}
\end{equation}
The equation (\ref{endpoint1}) also  implies that the
eigenvalues which are kept fixed  at the end points correspond
to the following operators
\begin{equation}
p_x(\tau'),\quad
p_t(\tau'),\quad\pi(\tau'),\quad{\cal C}(\tau'),\quad{\bar {\cal
C}}(\tau'),\label{BC11}
\end{equation}
\begin{equation}
x(\tau''),\quad t(\tau''),\quad
\pi(\tau''),\quad{\cal C}(\tau''),\quad{\bar {\cal C}}(\tau''),\label{BC12}
\end{equation}
which means that we are selecting  the following basis for the Hilbert space
\begin{equation}
\left\{|p_x',p_t',
\pi',{\cal C}',{\bar {\cal C}}',\tau'\rangle \equiv |\tau' \rangle \right\},
\quad\quad  \left\{\langle x'',t'',\pi'',{\cal C}'',{\bar {\cal C}}'',\tau''|
\equiv \langle \tau''|\right \},\label{basis1}
\end{equation}
at the initial  and  final  end-points respectively.
The  eigenvalues  $\pi',\pi'',{\cal C}',{\cal C}'', $ ${\bar {\cal C}}',
{\bar {\cal C}}''$ are taken to be zero, according to Eq.(\ref{cbc0}).
Our notation is $A''\ (A')$ for the eigenvalues
of the operator $A(\tau'') \ (A(\tau'))$. However, in order  to make the
notation not too cumbersome, we will denote with the same
letter, both the operator and its corresponding eigenvalue in the sequel,
expecting  that no confusion arises.

{}From the (anti)commutation relations (\ref{com1}) we can show that the
BRST operator $\Omega$ constructed in (\ref{addop1}) is hermitian and
nilpotent.
Also,
the BRST-invariance of the above basis (\ref{basis1})  can be directly
verified.
The calculation of the effective Hamiltonian can now be performed, leading
to
\begin{equation}
H_{BRST}= i{\bar {\cal P}}{\cal
P}+\lambda G,\label{effham1}
\end{equation}
which is  an hermitian operator satisfying
$ [ H_{BRST}, \Omega]= 0 $. The equations of motion
can  be written in  the following explicit form
\begin{eqnarray}
\dot p_x=0, & & \dot x -\frac{\lambda p_x}{m}=0,\quad
\dot p_t =0,\quad\dot t - \lambda=0,\quad
\dot\pi + G=0, \quad \dot\lambda=0, \nonumber \\
& & \dot{\cal P}=0,\quad \dot{\bar {\cal C}} -i{\bar {\cal P}}=0,\quad
\dot{\bar {\cal P}}=0,\quad \dot{\cal C}+i{\cal P}=0. \label{eqfp}
\end{eqnarray}
Next we consider the calculation of the propagator. The first step is
to solve the above operator equations. We obtain the general
solution
\begin{equation}
p_x=p'_x, \quad x(\tau)=x'+\frac{\lambda p_x}{m}(\tau-\tau'),\quad
p_t=p'_t, \quad t(\tau)=t'+\lambda(\tau-\tau'),\label{sol11}
\end{equation}
\begin{equation}
\pi(\tau)=\pi'-G (\tau-\tau'),\quad \lambda=\lambda',\label{sol12}
\end{equation}
\begin{equation}
{\cal P}={{\cal P}}', \quad{\bar {\cal C}}(\tau)={\bar {\cal C}}'+i{\bar {\cal
P}}(\tau-\tau'),\quad
{\bar {\cal P}}={{\bar {\cal P}}}', \quad {\cal C}(\tau)={\cal C}'-i{\cal
P}(\tau-\tau'),\label{sol13}
\end{equation}
where the superscript $'$  denotes the evaluation of the corresponding
operator at $\tau=\tau'$, which are used here to denote  arbitrary operator
integration constants to be further specified according to the boundary
conditions (\ref{basis1}). A slightly rewritten expression
for the variation for the propagator, obtained from (\ref{endpoint1}), is
\begin{eqnarray}
\delta\langle\tau''|\tau'\rangle=i\langle\tau''|p_x'\delta x''+x'\delta p_x'
+p'_t\delta t''+t'\delta p_t' -\lambda '(\delta\pi''-\delta\pi')\nonumber\\
-{\bar {\cal P}'}(\delta{\cal C}'' -\delta{\cal C}')-{\cal P}'(\delta{\bar
{\cal
C}}''-\delta{\bar {\cal C}}')-{H_{BRST}}(\delta\tau''
-\delta\tau')|\tau'\rangle.\label{var1}
\end{eqnarray}
The next step is to calculate the corresponding matrix elements.
After  we incorporate  the chosen boundary conditions (\ref{basis1}) in the
above solutions
(\ref{sol11})-(\ref{sol13})
of the equations of motion,  we can write $ H_{BRST}$
in terms of the end-points operators ${\cal C}',{\cal C}'',{\bar {\cal C}}',
{\bar {\cal C}}''$,  together with the constant operators  $\lambda$ and $G$.
The result is
\begin{equation}
H_{BRST}=\frac{i}{(\tau''-\tau')^2}\left({\bar {\cal C}}''{\cal C}''-{\bar
{\cal
C}}''{\cal C}'+{\cal C}''{\bar {\cal C}}'+{\bar {\cal C}}'{\cal C}'+
(\tau''-\tau')\right)+\lambda ' G  \label{hef}
\end{equation}
where the {\em well-ordering} ( $''$ operators to the left and $' $ operators
to the right) has been achieved by  using the anticommutator
\begin{equation}
\{{\bar {\cal C}}',{\cal C}''\}=-(\tau''-\tau'),\label{cbc}
\end{equation}
which is calculated from the solutions (\ref{sol11})-(\ref{sol13}),
together with the equal-time (anti)commutators (\ref{com1}).
The hermiticity of Eq. (\ref{hef})
can be verified  explicitly by
using again the relation (\ref{cbc}).

All the terms whose matrix elements produce eigenvalues that
are fixed to zero at boundaries do not contribute to
the propagator,  as it is the case of the ghosts and anti-ghosts.
Furthermore, reparametrization invariance demands that the
propagator be independent of the end-point values of the parameter
$\tau$. This is guaranteed provided that the matrix elements of  $H_{BRST}$
are zero.
In order to show this, we need to
calculate the matrix elements for $\lambda= {\lambda}'$.
This can be done as follows:
multiply from the left the first Eq.(\ref{sol12}) by $\lambda$ and take
the appropriate
matrix elements on both sides of the resulting equation. Then, use the fact
that the eigenvalues of $\pi$ are fixed to zero at the boundaries,
together with the equal-time commutator of $\lambda$ and $\pi$. The result
is
\begin{equation}
(\tau''-\tau')\langle\tau''|\lambda|\tau'\rangle=-{i\langle\tau''|\tau'
\rangle\over({{p_x}'}^{\ 2}/2m)+{p_t}'},\label{EMML1}
\end{equation}
which immediately implies  that
$\langle\tau''| H_{BRST} |\tau'\rangle=0 $.
As usual, we need to complete the  rewriting of  the variation (\ref{var1}) in
{\em well ordered form }.
In our case, this procedure has to be further  applied to the operators
 $x'$, ${p''}_t$ and $t'$. Using the corresponding
equations of motion we obtain
\begin{eqnarray}
\delta\langle\tau''|\tau'\rangle&=&i\langle\tau''| p_x'\delta x''+
(x''-\frac{\lambda' p_x'}{m}(\tau''-\tau'))\delta p_x'+p_t'\delta t''
\nonumber\\
&+&(t''-\lambda '(\tau''-\tau'))\delta p_t'|\tau'\rangle.
\end{eqnarray}
Finally, after substituting
the matrix elements of $\lambda$,  we are able to integrate
the resulting system of partial differential equations, obtaining
\begin{equation}
\langle x'',t'',\tau'' |p_x',p_t',\tau'\rangle= \exp\{ip_x'x''+
ip_t't''\}/[({p_x}'^{\ 2}/2m)+{p_t}'],
\end{equation}
which is the correct propagator for the  parametrized free particle.

An important point that we want to emphasize
is the
following :
suppose we have constructed  a  reparametrization invariant
version of an arbitrary theory defined through the Hamiltonian $H_0$, by
introducing
the parameter $\tau$ in complete analogy to the example considered in this
section. Under these
circumstances,
the extended Hamiltonian  will be always  proportional to the first-class
constraint
\begin{equation}
p_t+H_0(q,p)\approx 0,
\end{equation}
which arises as a consequence or the imposed reparametrization invariance.
The associated quantum condition upon the physical states is that
they must be annihilated by such constraint, which means that such states
can not depend on the parameter $\tau$ and, consequently, the
propagator must also be $\tau$ independent. In other words, the matrix
elements of the extended Hamiltonian between the physical states
must be zero.
The same argument is valid  for the matrix elements of the  BRST-Hamiltonian
between physical states, when we consider a
non-canonical fermionic gauge fixing   $\Psi=\bar{\cal P}\lambda$, in the BRST
approach for a reparametrization-invariant theory.
The latter property, which we have explicitly verified in the case of the
parametrized non-relativistic free particle,  is  in contradiction with the
results presented in
Ref. \cite{rabello}.

\section{The relativistic particle}

 Before considering this problem, let us  emphasize two important points
which
can be directly inferred from the previous example : (i) in the case where the
dynamics of the ghost-antighost sector
of the theory  decouples  from the remaining variables,
the effective Hamiltonian has the same form as in Eq.(\ref{hef}), except that
$G$ is now  replaced by the corresponding first-class constraint.
(ii)  we can always calculate the matrix elements of the
Lagrange multiplier associated with the reparametrization-invariance
constraint, by imposing the condition that the matrix elements of the
BRST-Hamiltonian are zero.

With this ideas in mind we now consider the calculation of the propagator for
the relativistic free particle from the point of view of the BRST-BFV operator
formulation. We start from the classical  action
\begin{equation}
S=\int_{\tau'}^{\tau''}d\tau \frac12 \left( {1 \over \lambda}{\dot x}^\mu
{\dot
x}_\mu-
\lambda m^2\right),
\end{equation}
which is reparametrization-invariant provided $\lambda$
transforms as a Lagrange multiplier.  Here we are taking the standard
Minkowski metric $\eta^{\mu\nu}=  \\ \mbox{diag}(-1,1,1,1)$.
The corresponding first-class constraint is now
\begin{equation}
\ G=p^\mu p_\mu + m^2\approx 0.
\end{equation}
Our starting point in the quantum problem is the operator effective action
\begin{eqnarray}
{S}_{BRST}&=&\int_{\tau'}^{\tau''}\left( <<{\dot{ x}}^\mu {p}_\mu -
{{ \lambda}}{\dot { \pi}} +
{\dot{{\bar {\cal C}}}} { {\cal P}}+{\dot{{\cal
C}}}{\bar {\cal
P}} >> - { H}_{BRST} \right)d\tau \nonumber \\
& & + <<{ x'}^\mu{{ p'}_\mu}>>,
\label{qacfp1}
\end{eqnarray}
with
\begin{equation}
{ H}_{BRST}=i \{ { \Psi }, \  { \Omega } \},\quad
{\Psi}= { {\bar {\cal P}}}{ \lambda}, \quad
{\Omega}=-i{{\cal P}}{ \pi} + {{\cal C}}\left (p^\mu p_\mu + m^2
\right),\label{addop}
\end{equation}
where the BRST-charge
has the same structure as in Eq.(\ref{addop1}) except for the
explicit form of the constraint $G$.
Here we are dropping the hats over the operators from the very beginning, in
order to simplify the notation. Starting from the action principle, in a
manner
completely analogous to the previous section, we obtain the following
non-zero
commutation relations
\begin{equation}
[x^\mu,p_\nu]=i\delta^\mu_\nu,\qquad [\lambda, \pi]=i,
\end{equation}
while the ghosts satisfy those anticommutators given in Eq.(\ref{com1}). The
(anti)\\ commutator algebra allows for the calculation of the
BRST-Hamiltonian
\begin{equation}
H_{BRST} = i{\bar {\cal
P}}{\cal P}+\lambda (p^2+m^2),
\end{equation}
together with the explicit form of the equations of motion
\begin{equation}
\dot p_\mu=0,\quad \dot{x^\mu} -2\lambda p^\mu=0,\quad
\dot \pi +G=0,\quad \dot\lambda=0,
\end{equation}
\begin{equation}
\dot{\cal P}=0,\quad \dot{\bar {\cal C}} -i{\bar {\cal P}}=0,\quad
\dot{\bar {\cal P}}=0,\quad \dot{\cal C}+i{\cal P}=0.
\end{equation}
The solution of the above equations is
\begin{equation}
p_\mu={p_\mu}' ,\quad {x^\mu}(\tau)={x^\mu}'+2\lambda p_\mu (\tau-\tau'),
\label{sol21}
\end{equation}
\begin{equation}
\pi(\tau)=\pi'-G(\tau-\tau'),\quad \lambda=\lambda',\label{sol22}
\end{equation}
\begin{equation}
{\bar {\cal P}}={{\bar {\cal P}}}', \quad {\bar {\cal C}}(\tau)={\bar {\cal
C}}'+i{\bar {\cal P}}(\tau-\tau'),\quad
{\cal P}={{\cal P}}', \quad {\cal C}(\tau)={\cal C}'-i{\cal
P}(\tau-\tau'),\label{sol23}
\end{equation}
where the primed operators  denote integrations constants to be determined
according to the
choice of the end-point conditions. The BRST-invariant boundary conditions are
chosen in complete analogy with
the previous section by fixing the operators
\begin{equation}
p_\mu(\tau'),\quad \pi(\tau'),\quad {\cal C}(\tau'), \quad {\bar {\cal
C}}(\tau'),\label{BC21}
\end{equation}
\begin{equation}
x^\mu(\tau''),\quad \pi(\tau''),\quad {\cal C}(\tau''),\quad {\bar {\cal
C}}(\tau''),
\label{BC22}
\end{equation}
at the end-points.
This choice implies  that the corresponding basis of the Hilbert space
are
\begin{equation}
\left\{|{p_\mu}',{\pi}',{\cal C}',{\bar {\cal
C}}',\tau'\rangle\right\}, \quad
\quad  \left\{\langle {x^\mu}'',{\pi}'',{\cal C}'',{\bar {\cal
C}}'',\tau''|\right\},
\end{equation}
respectively.  Again,   the eigenvalues
$\pi', \pi'', {\cal C}', {\cal C}'', {\bar {\cal C}}', {\bar {\cal C}}''$ are
taken  to be  zero in order to enforce the  BRST-invariance.
Since the effective Hamiltonian for this theory has the same structure
as in Eq.(\ref{hef}), we calculate the matrix elements for $\lambda$ by
demanding a null result for the matrix elements of $H_{BRST}$. The
answer is
\begin{equation}
(\tau''-\tau')\langle\tau''|\lambda|\tau'\rangle=-i{\langle\tau''|\tau'\rangle
\over p'^{\ 2}+m^2},
\end{equation}
which is analogous to that of Eq.(\ref{EMML1}).
Next we calculate the propagator. Its variation is given by
\begin{eqnarray}
\delta\langle\tau''|\tau'\rangle=i\langle\tau''|{p_\mu}'\delta {x^\mu}''+
{x^\mu}'\delta {p_\mu}'  -\lambda'(\delta {\pi}''-\delta
{\pi}')\nonumber\\
-{\bar {\cal P}}'(\delta{\cal C}''-\delta{\cal C}') -{\cal P}'
(\delta{\bar {\cal C}}'' -
\delta{\bar {\cal C}}')-{H_{BRST}}(\delta\tau''-
\delta\tau')|\tau'\rangle.
\end{eqnarray}
Using the solutions given in Eqs. (\ref{sol21})-(\ref{sol23})
written in terms of the operators fixed at the end-points, we obtain
\begin{equation}
\delta\langle\tau''|\tau'\rangle=i\langle\tau''| {p_\mu}'\delta {x^\mu}''+
({x^\mu}''-2\lambda p_\mu'(\tau''-\tau'))\delta {p_\mu}'|\tau'\rangle,
\end{equation}
in a manner completely similar to the previous case.
Finally, introducing the matrix elements of $\lambda$ and integrating with
respect to the end point eigenvalues,  we get the result
\begin{equation}
\langle {x^\mu}'',\tau'' |{p_\mu}',\tau'\rangle= \exp\{{ip_\mu}'{x^\mu}''\}
/[p'^{\ 2}+ m^2],
\end{equation}
which gives the  propagator for the free relativistic particle.

\section{The spinning relativistic free particle}

As our final example we consider the spinning relativistic free particle.
To this end let us start from the following classical action
\begin{equation}
S=\int_{\tau'}^{\tau''}d\tau (\dot{x^\mu}p_\mu +\frac{i}{2}(\dot{\theta^\mu}
\theta_\mu+ \dot{\theta_5}\theta_5)-
N{\cal H} -M {\cal Q}_0)-\frac{i}{2}\theta(\tau'')\cdot\theta(\tau')
-[B]^{\tau''}_{\tau'},\label{action3}
\end{equation}
where the variables  $x^\mu, p_\mu, N, {\cal H}$ are real-even
Gras\-smann-valued,
while $\theta^\mu, \theta_5,\ { {\cal Q}_0}$ are correspondingly real-odd and
$M$ is imaginary-odd, in accordance with our general conventions.
The first class constraints ${\cal H}$ and ${ {\cal Q}_0}$ are
\begin{equation}
{\cal H}=p^\mu p_\mu+m^2,\quad { {\cal Q}_0}=p_\mu\theta^\mu + m\theta_5.
\end{equation}
The explicitly written boundary term $-\frac{i}{2}\theta(\tau'')
\cdot\theta(\tau')=-{i\over2}(\theta^\mu(\tau'')
\theta_\mu(\tau')+\theta_5(\tau'')\theta_5(\tau')) $ provides
the correct end-point conditions for the fermionic coordinates $\theta^\mu,
\theta_5$
leading to the fixing of the following combinations \cite{henneaux}
\begin{equation}
\frac12(\theta^\mu(\tau')+\theta^\mu(\tau''))\equiv\xi^\mu, \quad
\frac12(\theta_5(\tau')+\theta_5(\tau''))\equiv\xi_5, \label {FBC}
\end{equation}
which provide unique solutions to the corresponding first-order
equations of motion. There could still be additional boundary terms in the
action (\ref{action3}), related to the choice of the end points  conditions
for the remaining
variables, which are contained in $B$.
Next we go through the classical  BRST formalism. Let us introduce
the
vector
\begin{equation}
G_A=(\pi_b, G_b)=(\pi_M,\pi_N,{ {\cal Q}_0},{\cal H}),\quad b=1,2,
\end{equation}
where the new variables $\pi_M$ and $\pi_N$ are the momenta canonically
conjugated to the Lagrange multipliers $M$ and $N$.
Here $\epsilon_1=1, \epsilon_2=0$.  The ghosts and
anti-ghosts
are taken to be  \begin{equation}
\eta^A=(-{\cal P}^1,-i{\cal P}^2,{\cal C}^1,{\cal C}^2), \qquad
{\cal P}_A=({\bar {\cal C}}_1, i{\bar {\cal C}}_2,{\bar {\cal P}}_1,{\bar
{\cal P}}_2),
\end{equation}
where $({\cal P}^1,{\bar {\cal C}}_1), ({\cal C}^1,{\bar {\cal P}}_1)$ are
even canonically-conjugated
ghost-antighost variables while $({\cal P}^2,{\bar {\cal C}}_2), ({\cal
C}^2,{\bar {\cal P}}_2)$ are correspondingly
odd .
With these ingredients we now construct the classical  BRST charge.
The general expression for the case under consideration is
\begin{equation}
\Omega=\eta^AG_A-\frac{1}{2}(-1)^{\epsilon_B}\eta^B\eta^CC_{CB}^A{\cal P}_A,
\label{OM}
\end{equation}
where $\epsilon_B$ is the Grassmann parity of the constraint
associated with the
variable $b$
and $C_{CB}^A$ are
the structure functions of the algebra of constraints, which in this case is
given by
\begin{equation}
\{{ {\cal Q}_0},{ {\cal Q}_0}\}_{PB}=i{\cal H}, \qquad \{{ {\cal Q}_0},{\cal
H}\}_{PB}=0.
\end{equation}
Accordingly,
the only structure function different from zero is
$C_{11}^2=i$. Taking this into account and making the required
substitutions in Eq.(\ref{OM}) we get
\begin{equation}
\Omega=-{\cal P}^1\pi_M-i{\cal P}^2\pi_N+{\cal C}^1{ {\cal Q}_0}+{\cal
C}^2{\cal
H}+i({\cal C}^1)^2{\bar {\cal P}}_2,
\end{equation}
for the classical BRST charge.
The theory  considered in this section is also reparametrization
invariant and thus the canonical Hamiltonian is zero.

Now we promote all dynamical variables to operators with the following reality
properties:
 $x_\mu,
p^\mu, N$, $\pi_N, {\bar {\cal P}}_1$, ${\cal C}^1, {\cal P}^1, {\bar {\cal
C}}_1$ are hermitian-even
operators, ${\cal P}^2, {\bar {\cal P}}_2, M$ are antihermitian-odd operators
and
${\bar {\cal C}}_2,{\cal C}^2, \theta^\mu, \theta_5, \pi_M$ are hermitian-odd
operators. The quantum effective action that we start from  is
\begin{eqnarray}
S_{BRST}&=&\int^{\tau''}_{\tau'} d\tau \left( << \dot{x^\mu}{p_\mu}
+\frac{i}{2}(\dot{\theta^\mu}
\theta_\mu+\dot{\theta_5}\theta_5)- N{\dot\pi}_N-
 M{\dot\pi}_M  \right. \nonumber \\
&  & + \left.   {\dot { \bar {\cal C}}}_1{\cal P}^1 +\dot {{\cal C}^1} {\bar
{\cal P}}_1
+{\dot {\bar {\cal C}}}_2{\cal P}^2 +\dot {{\cal C}^2} {\bar {\cal P}}_2 >>
-H_{BRST} \right) \nonumber \\
& &  + <<- \frac{i}{2}\theta(\tau'')
\cdot\theta(\tau') +{x^\mu}' {p_\mu}'>>, \label{VP}
\end{eqnarray}
where
\begin{eqnarray}
& & { H}_{BRST}=i \{ { \Psi }, \  { \Omega } \},\quad
{\Psi}= <<{\bar {\cal P}}_1M+{\bar {\cal P}}_2N>>,  \nonumber \\
& & \Omega=<<-{\cal P}^1\pi_M-i{\cal P}^2\pi_N+{\cal C}^1{ {\cal Q}_0}+{\cal
C}^2{\cal
H}+i({\cal C}^1)^2{\bar {\cal P}}_2>>.
\end{eqnarray}
The (anti)commutation relations arising from the action principle are
\begin{equation}
[x_\mu,p_\nu]=i\eta_{\mu\nu} \quad
\{\theta^\mu,\theta^\nu\}=-\eta^{\mu\nu}\quad
\{\theta_5,\theta_5\}=-1,\label{7}\\
\end{equation}
\begin{equation}
\{M,\pi_M\}=-i\quad [N,\pi_N]=i,\nonumber\\
\end{equation}
\begin{equation}
[{\bar {\cal P}}_1,{\cal C}^1]=-[{\cal C}^1,{\bar {\cal P}}_1]=[{\cal
P}^1,{\bar {\cal C}}_1]=-[{\bar {\cal C}}_1,{\cal P}^1]=-i,\nonumber\\
\end{equation}
\begin{equation}
\{{\bar {\cal P}}_2,{\cal C}^2\}=\{{\cal C}^2,{\bar {\cal P}}_2\}=\{{\bar
{\cal
C}}_2,{\cal P}^2\}=\{{\cal P}^2,{\bar {\cal C}}_2\}=-i. \label{CR}
\end{equation}
The calculation of the fermionic anticommutators in Eq.(\ref{7}) is a
particular case of the work in Ref.\cite{das}.
Using the above results one can directly verify
the anticommutator $\{ \Omega, \ \Omega
\}=0 $ and also we can calculate the BRST- Hamiltonian
\begin{equation}
H_{BRST}=-
{\bar {\cal P}}_1{\cal P}^1+M{ {\cal Q}_0}+2iM{\cal C}^1{\bar {\cal
P}}_2+i{\bar {\cal P}}_2{\cal P}^2+N{\cal H}, \label{hsrp}
\end{equation}
which leads to the following explicit form for the quantum effective action
\begin{eqnarray}
S_{BRST}&=& \int^{\tau''}_{\tau'} d\tau \left( << \dot{x^\mu}{p_\mu}
+\frac{i}{2}(\dot{\theta^\mu}
\theta_\mu+\dot{\theta_5}\theta_5)- N{\dot\pi}_N-
 M{\dot\pi}_M \right. \nonumber\\
& & \left. +{\dot{\bar {\cal C}}}_1{\cal P}^1+\dot{{\cal
C}^1}{\bar {\cal P}}_1
+{\dot{\bar {\cal C}}}_2 {\cal P}^2+\dot{{\cal C}^2}{\bar {\cal P}}_2 >>
+{\bar {\cal P}}_1{\cal P}^1-M{ {\cal Q}_0} \right. \nonumber \\
& & \left. -2iM{\cal C}^1{\bar
{\cal P}}_2-i{\bar {\cal P}}_2{\cal P}^2-N{\cal H} \right) \nonumber \\
& & << -
\frac{i}{2}\theta(\tau'')
\cdot\theta(\tau') +{x^\mu}' {p_\mu}'>>.
\end{eqnarray}
The reality properties of the remaining operators are:
${\cal H}, H_{BRST}$ are hermitian-even,  ${ {\cal Q}_0}, \Omega$ are
hermitian-odd, while $\Psi$ is antihermitian-odd.
The corresponding equations of motion are
$$
\dot{p^\mu}=0,\quad \dot{x^\mu}-M\theta^\mu-2Np^\mu=0,\quad
\dot{\theta^\mu}+iMp^\mu=0,\quad \dot{\theta_5}+iMm=0,
$$
$$
\dot{\pi}_M+{ {\cal Q}_0}+2i{\cal C}^1{\bar {\cal P}}_2=0,\quad \dot M=0,\quad
\dot{\pi_N}+{\cal H}=0,\quad\dot N=0,
$$
$$
\dot{{\cal P}^1}=0,\quad\dot{{\bar {\cal C}}_1}+{\bar {\cal P}}_1=0,\quad
\dot{{\cal C}^1}+{\cal P}^1=0,\quad \dot{{\bar {\cal P}}_1}+2iM{\bar {\cal
P}}_2=0,
$$
$$
\dot{{\cal P}^2}=0,\quad\dot{{\bar {\cal C}}_2}-i{\bar {\cal P}}_2=0,\quad
\dot{{\bar {\cal P}}_2}=0,\quad \dot{{\cal C}^2}-2iM{\cal C}^1+i{\cal P}^2=0.
$$
The general solution of the above system is
\begin{equation}
p_\mu={p_\mu}' \ ,{\cal P}^1={{\cal P}^1}' \ ,M=M' \ ,N=N' \ ,{\cal
P}^2={{\cal
P}^2}', \ {\bar {\cal P}}_2=
{{\bar {\cal P}}_2}',
\end{equation}
\begin{equation}
x^\mu(\tau)=x'^\mu+(M\xi^\mu+2Np^\mu)(\tau-\tau'),
\end{equation}
\begin{equation}
\theta^\mu(\tau)=-iMp^\mu\tau+\xi^\mu+\frac{i}{2}Mp^\mu(\tau''+\tau'), \label
{TMU}
\end{equation}
\begin{equation}
\theta_5(\tau)=-iMm\tau+\xi_5+\frac{i}{2}Mm(\tau''+\tau'),
\end{equation}
\begin{equation}
{\cal C}^1(\tau)={\cal C}'^1-{\cal P}^1(\tau-\tau'),
\end{equation}
\begin{equation}
{\bar {\cal P}}_1(\tau)={\bar {\cal P}}'_1-2iM{\bar {\cal P}}_2(\tau-\tau'),
\end{equation}
\begin{equation}
{\bar {\cal C}}_1(\tau)={\bar {\cal C}}'_1-({\bar {\cal P}}'_1-iM{\bar {\cal
P}}_2(\tau-\tau'))(\tau-\tau'),
\end{equation}
\begin{equation}
{\bar {\cal C}}_2(\tau)={\bar {\cal C}}'_2+i{\bar {\cal P}}_2(\tau-\tau'),
\end{equation}
\begin{equation}
{\cal C}^2(\tau)={\cal C}'^2+iM(2{\cal C}'^1-{\cal
P}^1(\tau-\tau'))(\tau-\tau')-i
{\cal P}^2(\tau-\tau'),
\end{equation}
\begin{equation}
\pi_N(\tau)=\pi'_N-{\cal H}(\tau-\tau'),\label{PIN}
\end{equation}
\begin{equation}
\pi_M(\tau)=\pi'_M-({ {\cal Q}_0}+i(2{\cal C}'^1-{\cal P}^1(\tau-\tau')){\bar
{\cal P}}_2)
(\tau-\tau').\label{PIM}
\end{equation}
The notation is the same as in the previous sections.

The boundary conditions that we take  are completely
similar to our  previous cases.
The only novelty  that we encounter  here is related to the fermionic
degrees of freedom described by the $\theta$-variables. In order to
clearly elucidate this point, let us consider for a moment the
contribution of the fermionic degrees of freedom
$\theta^\mu$ to the change of the effective action
\begin{eqnarray}
\delta_{\theta_\mu} {S_{BRST}} & =
&\int_{\tau'}^{\tau''}d\tau(\frac{i}{2}\dot{\theta^\mu}
\delta\theta_\mu +\frac{i}{2}\delta\dot{\theta^\mu}\theta_\mu-M\delta
\theta_\mu p^\mu) \nonumber \\ & -
&\frac{i}{2}(\delta\theta^\mu(\tau'')\theta_\mu(\tau')+
\theta^\mu(\tau'')\delta\theta_\mu(\tau'))\nonumber\\
& = &\int_{\tau'}^{\tau''}d\tau((i\dot{\theta^\mu}-Mp^\mu)\delta\theta_\mu)+
\frac{i}{2}\delta(\theta^\mu(\tau')+\theta^\mu(\tau''))(\theta_\mu(\tau'')-
\theta_\mu(\tau')).\nonumber \\
\end{eqnarray}
Substituting the solution of the equations of motion for $\theta_\mu$
(\ref{TMU}),
together with  the definition (\ref{FBC}) of the variable $\xi^\mu$
we obtain
\begin{equation}
\delta_{\theta_\mu} {S_{BRST}}=\delta\xi^\mu M p_\mu (\tau''-\tau').
\end{equation}
The same analysis  can be applied to $\theta_5$.

The end-point operators are chosen in such a way that the
fixed eigenvalues are
\begin{equation}
p_\mu(\tau')={p_\mu}',\qquad x^\mu(\tau'')={x^\mu}''
\end{equation}
\begin{equation}
\frac12(\theta^\mu(\tau')+\theta^\mu(\tau''))=\xi^\mu,\quad
\frac12(\theta_5(\tau')+\theta_5(\tau''))=\xi_5,
\end{equation}
\begin{equation}
\pi_N(\tau')=\pi_N(\tau'')={\cal C}^i(\tau')={\cal C}^i(\tau'')={\bar {\cal
C}}_i(\tau')=
{\bar {\cal C}}_i(\tau'')=0, \quad i=1,2,
\end{equation}
together with the corresponding BRST-invariant basis
\begin{equation}
\{ \langle {x^\mu}'',{\pi_N}'',{\pi_M}'', {\theta''}^\mu(\xi^\mu),
{\theta''}_5(\xi_5),
{{\cal C}^i}'',{{\bar {\cal C}}_i}'',\tau''| \},$$
$$ \{ |{p_\mu}',{\pi_N}', {\pi_M}',{\theta'}^\mu(\xi^\mu),
{\theta'}_5(\xi_5),
{{\cal C}^i}',{{\bar {\cal C}}_i}',\tau'\rangle\}.
\end{equation}
Before going to the calculation of the propagator, let us rewrite the
effective Hamiltonian (\ref{hsrp}) in well ordered form. To this end, we use
the
equations of motion
together with the following (anti)commutation relations at different times
\begin{equation}
[{\bar {\cal C}}_1',{{\cal C}^1}'']=-i(\tau''-\tau'),
\quad \{{\bar {\cal C}}_2',{{\cal C}^2}''\}=-(\tau''-\tau').
\end{equation}
The result is
\begin{eqnarray}
H_{BRST}&=&\frac{1}{(\tau''-\tau')^2}[-({\bar {\cal C}}_1''{{\cal
C}^1}''-{\bar {\cal C}}_1''
{{\cal C}^1}'-{{\cal C}^1}''{\bar {\cal C}}_1'+{\bar {\cal
C}}_1'{{\cal C}^1}')\nonumber\\
&+&i({\bar {\cal C}}_2''{{\cal C}^2}''-{\bar {\cal C}}_2''{{\cal
C}^2}'+{{\cal C}^2}''{\bar {\cal C}}_2'+
{\bar {\cal C}}_2'{{\cal C}^2}')]\nonumber\\
&+&M(\xi^\mu p'_\mu+m\xi_5)+N(p'^2+m^2),
\end{eqnarray}
Again, it is a
direct matter to verify that $H_{BRST}$ is hermitian.

Before going to the calculation of the propagator it is necessary
to establish the following results
\begin{equation}
(\tau''-\tau')\langle\tau''|N|\tau'\rangle{\cal
H}'=-i\langle\tau''|\tau'\rangle,\quad
(\tau''-\tau')\langle\tau''|M|\tau'\rangle{ {\cal Q}'_0}=
i\langle\tau''|\tau'\rangle,
\end{equation}
where ${\cal H}'=(p'^2+m^2)$ and ${\cal Q}'_0=\xi^\mu p'_\mu+m\xi_5$.
These matrix elements
are calculated from Eqs.(\ref{PIN}) and (\ref{PIM}) respectively
and again imply the condition that the matrix
elements of the BRST-Hamiltonian between physical states
must be zero.

The general variation of the propagator is
\begin{equation}
\delta\langle\tau''|\tau'\rangle=i\langle\tau''|p'_\mu\delta {x^\mu}''+
{x^\mu}'\delta {p_\mu}'  +\delta\xi^\mu Mp'^\mu(\tau''-\tau') +
\delta\xi_5 Mm(\tau''-\tau')
|\tau'\rangle.
\end{equation}
After substituting the solution of the equations of motion
for $x'^\mu$ in terms of the boundary operators and after performing the
necessary integrations,
we obtain the required propagator
\begin{equation}
\langle \tau''|\tau'\rangle=\exp[ip'x'']{(p'_\mu\xi^\mu + m\xi_5)\over
(p'^2+m^2 ) }.
\end{equation}

In conclusion,  starting from the  quantized version of the BRST-BFV effective
action given in Eq.(\ref{qeffac}),
together with the use of  the  Schwinger action principle and the
imposition of correct BRST-invariant
boundary conditions, we have obtained the
propagators of the parametrized non-relativistic free particle and of  the
relativistic free particle, in the  spinless and spining cases.
\section{ Acknowledgements}
LFU, JDV and JAG were partially supported by the grant UNAM-DAGAPA-IN100694.
LFU and JDV also ac\-knowledge support from the grant
CO\-NA\-CyT-400200-5-3544E. JAG is supported by the  CONACyT  graduate
fellowship \# 86226.

\end{document}